# Why Aren't More Theories Named After Women?: Teaching Women's History in Physics

Beth Parks, Colgate University, meparks@colgate.edu

**Abstract**
Barriers to women's education and employment in Europe and the U.S. in the nineteenth century made it unlikely that any women would be among the few physicists whose ideas are taught in high school and college courses. This paper explores the social settings in which three influential physicists worked—James Clerk Maxwell, Robert Millikan, and Albert Einstein—to better understand the limited opportunities available to women. By acknowledging and explaining why there weren't more women among these founding physicists, instructors may help students understand the barriers that still exist and feel more empowered to overcome them and pursue physics as a career.

**Introduction**
Much effort has been expended to uncover forgotten contributions by women in physics and publicize examples of female physicists to students, largely to help female students see themselves in physics.[1-11] While these examples show that women can contribute to physics, the physicists who are credited with formulating the ideas taught in most high school and college courses were almost all men. The take-away message can easily be that an extraordinary woman may occasionally contribute to physics, but men have done foundational work. In my experience both as a learner and a teacher of physics, when this disparity goes unnoted, it is disheartening. If scouring the historical records to uncover hidden contributions of women still shows women to be unequal to men in physics, then what is implied about the possibility of women's success in the future? In my opinion, if we don't acknowledge and address the absence of women from these founding roles, female students won't feel fully supported in their decision to become physicists.

Instead, I believe we should ensure that students are aware of the social context in which these most influential physicists worked, so that students will see why it's not surprising that almost all were male. We assume that women and men have the same innate potential to succeed in physics, but developing this potential requires a combination of educational opportunities, family support, social acceptance, financial means, and employment opportunities. If these resources were available to a very much smaller fraction of women than men, then it's no surprise that a small fraction of foundational discoveries were made by women. To fully understand this disparity, in addition to looking at the lives of women who *did* become physicists, I suggest that we need to look at why so many *didn't*. In this article I'll present a brief sampling, looking at the educations and opportunities of three important male physicists—James Clerk Maxwell, Robert Millikan, and Albert Einstein—in different countries and times, and comparing them to their female peers.

**The U.K. in the mid-19th century—James Clerk Maxwell**
James Clerk Maxwell was born in Scotland in 1831. He was educated at home by his mother until age 8, then at age 10 entered Edinburgh Academy. Consider how opportunities in primary education differed for Maxwell's female peers. During this early Victorian period, middle-class boys were typically sent to school, while girls were educated at home. "Until the 1870s those who could afford it employed governesses or sent their daughters to small private schools . . . Schooling was considered a way for girls to obtain social rather than intellectual skills."[12] While there were some secondary schools for girls in Scotland, called "young ladies' institutions," these gave limited instruction in science and mathematics.[13] Girls were expected to learn to play an instrument, sing, knit, and sew. A Royal Commission report in 1864 concluded that girls' schools were characterized by "want of thoroughness and foundation; want of system; slovenliness and showy superficiality; inattention to rudiments; undue time given to accomplishments. . . ."[14]

While a few girls still might—perhaps if educated at home by liberal fathers—have been able to receive a secondary education that included algebra, geometry, and physics, the next steps in Maxwell's career were completely closed to his female peers. He entered Edinburgh University at age 16, then Cambridge University in 1850, at age 19. These institutions were not open to women until 1869. Those few women's colleges that did exist were designed to educate governesses, and did not include advanced courses in mathematics and physics. In the highly class-conscious Victorian period, upper- and middle-class girls obtained their status first through their fathers and then through their husbands. If they held a job, they decreased the status of their families. Unless a young woman needed to support herself as a governess, there was no thought of educating her for a career. Further, while Maxwell did his most important work while employed as a professor at Cambridge, Aberdeen, and King's College, the first appointment of a woman to a professorship in the U.K. was not until 1908. Those women who made contributions to physics in the U.K., such as Hertha Ayrton, were not able to support themselves by doing this work.

**The U.S. in the late 19th century—Robert Millikan**
Switching continents, we can consider the educational options open to American physicist Robert Millikan and his female peers. Millikan was born in 1868 in Illinois, attended Oberlin College for an undergraduate and master's degree, then in 1893 received a fellowship from Columbia University to study for a PhD. His most important work (determining the charge of the electron and measuring *h* via the photoelectric effect) was done while a faculty member at the University of Chicago, in a position starting in 1896.

At this time in the U.S., education was open to girls, who could study the same subjects in secondary school as boys and even continue to post-secondary education; in fact, Millikan's alma mater, Oberlin College, was co-educational. While there were certainly fewer opportunities for women, in 1890, 36% of college students were women.[15]

The difference between men's and women's opportunities occurred at the post-college level. In the post-civil-war period, for the most part, professions outside of pre-college teaching were closed to women. The high-profile women's colleges that opened during this period, Vassar, Smith, Wellesley, Radcliffe, and Bryn Mawr, aimed "to produce Christian women better prepared to assume their duties in the domestic sphere, as wives and mothers, and only if need be, as schoolteachers."[15] However, as the century drew to its end, women started to enter

Ph.D. programs, and by 1900, 228 women had earned doctorate degrees (2,372 men had done the same). One reason for this factor-of-ten difference was the difficulty for women in obtaining fellowships. Columbia University, where Millikan earned his PhD, reserved its most generous fellowships for men ($650, of which $600 was required to cover a year's tuition), and "out of thirty-two scholarships of $150, women could apply for only four."[15]

For those women who could manage the finances to earn a Ph.D., there was little open to them beyond the degree. As Margaret Rossiter explained, their careers were stymied by the movement to "professionalize" science, where this new concept of professionalization was inextricably linked to masculinity. Women were allowed to teach at women's colleges (where research support was minimal at best). Other roles in science were either professionalized and masculinized or downgraded and feminized. Women were hired to be "computers" and "assistants" despite possessing the same Ph.D. credentials as men hired to be professors and researchers. The existence of separate spheres for men and women was so accepted that job advertisements for scientists in the U.S. Civil Service were required to be written explicitly for either men or women.[16]

A final barrier was the incompatibility of marriage with education or employment for women. Only 16 of the women who earned PhDs in the nineteenth century were married at the time the degree was awarded, and married women were not eligible for fellowships. This incompatibility was so ingrained that the president of Vassar evidently thought he was expressing a particularly liberal viewpoint by stating, "if any love literature or art better than married life, that woman should be free to choose."[15] Faculty positions at women's colleges, which were the main options open to women Ph.D.s, were open only to unmarried women, and "any women already on the college faculty who did marry, as happened on occasion, resigned immediately."[16] In contrast, Millikan married in 1902 without incident.

**Germany at the turn of the century—Albert Einstein**
Finally, consider perhaps the most famous physicist of modern times, Albert Einstein. Einstein was born in 1879 in Ulm, Germany, and entered primary school in Munich in 1885. He left school in 1894 without a degree. He studied for the "school-leaving" certificate, which he passed in 1896 and entered university in Zurich with the goal of becoming a teacher, then earned a living through teaching, tutoring, and, famously, as a patent clerk, before submitting his PhD dissertation in 1905.

Einstein succeeded despite an unusual educational path, but a German girl living at the same time had no choice but to take an unusual educational path. Although Einstein was born a decade after Millikan, education for girls in Germany lagged considerably behind that in the U.S. In the U.S. at this time, secondary education was normally co-educational and nearly equal for girls and boys (and, in fact, girls graduated at a higher rate because boys left for employment). In contrast, as described by James Albisetti, German girls' education was still separate from boys'; did not prepare them for university; and was particularly lacking in science and mathematics. A curriculum issued in 1885 for one of the most rigorous girls' schools included two hours per week of arithmetic and science, with no higher mathematics. Women were explicitly banned from German universities from 1879. (The ban was lifted at individual universities beginning in 1900 and was not fully lifted until 1918.) If they wanted to prepare for entrance to Swiss universities, they had to enroll in private supplemental courses. While it was certainly possible for a woman to develop an interest in science or mathematics on her own and pursue it outside of normal secondary education, it is also fairly certain that many women never had the opportunity to discover or pursue this interest, or internalized societal pressures telling them that they did not belong. "As late as 1959, women composed less than 2 percent of the faculty of West German universities, and a survey revealed that 79 percent of male professors did not think that there should be any women in their ranks." Additionally, as in the U.S., a woman had to choose between marriage and pursuing or using her education; the ban on married female school teachers was not lifted until 1907, and, even then, only for women in exceptional circumstances.[17]

**How did any women succeed?**
These examples are not intended to say that it was *impossible* for women to contribute to physics, but just that the odds were so stacked against them that it is unsurprising that so few were able to make major contributions to physics research. (The situation was easier in fields such as astronomy and botany, where the contributions of amateurs were respected, and where observational skills could replace academic credentials and access to a laboratory.)[15]

In the biographies of those women physicists who did succeed, we see the simultaneous presence of several unusual advantages. Marie Curie's family supported girls' education, although that support could only provide for primary education. She studied on her own to prepare for university; worked as a governess to earn money; and then traveled to France to study, where she lived in extreme poverty, in an attic room without heat, lighting, or water, studying constantly and nearly starving. Her Ph.D. research was possible only because she had married a physicist, Pierre Curie, and was able to use a shed outside his institute to purify radium. Her father-in-law, Eugène Curie, cared for her daughter Irène, allowing her to continue her research following her daughter's birth. And while there is now no question of her contribution to the research, she was not nominated for the 1903 Nobel Prize, and only won it because her husband, Pierre, requested that she not be omitted. Even after receiving the Nobel Prize, she was not awarded a faculty position at the Sorbonne until she was appointed to Pierre's position following his death in 1906.[18]

Family connections were also important to Emmy Noether; her PhD advisor was her father's friend and colleague. She worked without compensation following her PhD in 1908 (presumably supported by her family), serving as a dissertation adviser to many students and publishing important work, and was not appointed to a salaried position until 1923.[19]

Lise Meitner was lucky enough to have been born in Vienna. The University of Vienna was one of a few institutions that, through the dedication of leaders, developed cultures more welcoming to women.[20] She was in one of the early groups of women allowed to take its entrance exam, which she prepared for with a private tutor paid by her family. She received her PhD in 1906, and in 1907 she traveled to Berlin and became affiliated with the Friedrich-

Wilhelm-Universität. It was not at that time open to women, but, as a compromise, she was permitted to set up a lab in a basement room (a former wood shed) with a separate entrance. After five years of working in Berlin, during which time she published more than 20 papers, and reached the age of 34, she still had no position and was supported by her family. Finally in 1912 she received an appointment to grade Max Planck's students' papers, and in 1913 became a "scientific associate" at her institute.[21]

**How have things changed?**
Histories such as these can explain why women are so under-represented in founding roles in physics. But presenting this history as an explanation for the dearth of women in the past will surely lead to another question—why are there so few women *now*? That, of course, is the subject of a different paper (or, rather, it is the subject of *many* different papers). The most egregious and explicit limitations on women are gone—women can study at nearly all institutions of higher learning, and discrimination in hiring is illegal in most countries.

But it may be interesting to discuss with students the echoes of these problems from the past that still linger today. One amusing thread is the presence (or absence) of women's toilets. In describing the construction of the Finsbury Technical College in London in 1883, Claire Jones writes, "In keeping with this masculine, industrial arrangement, the original plans had omitted any female toilets – an oversight that had to be remedied hastily when a need was recognized."[22] In contrast, one piece of evidence that the Radium Institute in Vienna (where Lise Meitner worked briefly) was particularly enlightened with regards to women was the presence of women's toilets.[20] Even today, it can be challenging to find a women's restroom in some university science buildings; as an external reviewer of a physics department in 2013 I had to explain that it was unacceptable to have a men's room on the entrance level without a women's room.

We've also seen that the conflict between marriage, child rearing, and physics research made careers in physics difficult in the 19th century. Women are no longer fired for marrying, but nepotism rules "continued throughout the twentieth century in some American universities and many still continue in other countries. . . . Unmarried women scientists with tenure could be—and were—removed by this rule if they later married."[23] Assumptions about marriage hurt in other ways. Helen Quinn, a pre-eminent particle physics theorist and a former president of the American Physical Society, said that her advisor, while encouraging her to go to graduate school, told her, *"*graduate schools are usually reluctant to accept women because they get married and they don't finish. But I don't think we need to worry about that with you." (She wasn't sure whether he meant she wouldn't marry or she would finish regardless.)[24] While no policies are currently this explicit, marriage and child rearing are still more challenging for female physicists than for males. For example, the 2010 IUPAP survey of 15,000 physicists from 30 countries found that, "women are more likely than men to report that they do more of the housework than their spouses or partners. That result holds even . . . if we consider only households in which the woman makes more than her partner." Additionally, "by an almost two-to-one margin, women were more likely than men to say that becoming a parent significantly affected their work in various ways."[25] Interviews can be even more enlightening.

A recent study asked 74 male physicists and biologists about work and family roles. One respondent concisely summarized the problem for women. When asked whether having children poses challenges for scientists, he responded, "No, absolutely not. That's why you have a wife."[26]

Finally, the opinions expressed about women's ability to do science in the 19th century seem either quaint or horrifying today. Dr. Henry Maudsley wrote, "Their nerve-centres being in a state of greater instability, by reason of the development of their reproductive functions, they will be the more easily and the more seriously deranged."[27] Although this argument now seems ludicrous, his writings greatly influenced the organization of the women's colleges in Cambridge.[22] The Darwinian arguments that female brains were not organized to do science are reflected in experiences of female students in the present day, such as, "being told: 'guys are better at this so you should do this (other thing).'"[28] Or, as Nancy Hopkins of MIT wrote in an online forum: "I have found that even when women win the Nobel Prize, someone is bound to tell me they did not deserve it, or the discovery was really made by a man, or the important result was made by a man, or the woman really isn't that smart. This is what discrimination looks like in 2011."[29]

**Recommendations and Conclusions**
The lack of opportunity discussed in these cases may seem obvious to faculty "of a certain age," who remember when the Ivies became co-ed and grew up in a time when the main professions available to women were teaching and nursing. But it may not be obvious to current students, whose great-great-great-grandmothers were contemporaries of the pioneering women who first entered universities. Some may still have lingering questions about women's ability to do physics (as is shown by the comments of male classmates). Therefore, I suggest that instructors find ways to incorporate some of this material into their classrooms.

Different approaches will be appropriate in different classrooms. Some instructors might already set aside some class time to discuss gender issues and could add a summary of this material; others might simply post this article on a class website. I teach a modern physics class that emphasizes the historical development of the wave-particle duality; in this class, after discussing the discoveries of the fourth or fifth white male physicist, I can take some time to discuss the reasons why the founders of physics weren't more diverse. (Of course, the opportunities available to racial and ethnic minorities were even more circumscribed; this would be the subject of a separate article.) In contrast, some instructors minimize the association of names with physics theories, arguing that the "great man or woman of science" language distorts the actual process of science. In such a classroom, the instructor might discuss how specific circumstances had to align to create this apparent success, and that both men and women found other ways to contribute to science through less prestigious roles in laboratories, as textbook authors, and as makers of scientific instruments.

Without a careful study of the effect of these lessons, I can't make definite recommendations on how to place them. It might be wise not to teach them immediately before an exam, in order to avoid activating stereotype threat.[30-32] I encourage instructors to try various approaches and to attempt to quantify their effect.  I hope that these ideas can serve as a tool to increase equity for women in physics.  By acknowledging the past, we can hope to move forward.

**Acknowledgments**

The author thanks Margaret Rossiter for introducing her to the scholarship and literature of the history of education.